\documentstyle[prb,aps,twocolumn]{revtex}

\def\vec#1{{\bf #1}}
\def\hatn#1{{\bf{\hat #1}}}

\begin{document}

\pagestyle{empty}

\narrowtext

{\bf Comment on ``Evidence for Anisotropic State of Two-Dimensional
  Electrons in High Landau Levels''}

In a recent letter M. Lilly et al\cite{Lilly} have shown that a highly
anisotropic state can arise in certain 2D electron
systems.  (This effect has also now been seen by another
group\cite{Du}).  Most of the samples studied are square. In these
samples, the resistances in the two perpendicular directions are found
to have a ratio $R_{yy}/R_{xx}$ that may be as much as 60 or even
larger at low temperature and at certain magnetic fields.  In the same
work, Hall bar measurements were also performed which can be thought
of, at least conceptually, as being measurements on rectangular
samples (with side lengths $L_x$ and $L_y$).  In such measurements
$R_{yy}$ is measured in a sample with $L_y \gg L_x$ and $R_{xx}$ in a
sample where $L_x \gg L_y$.  In the Hall bar experiments it is found
that the anisotropy ratio is much smaller --- with $R_{yy}/R_{xx}
\approx 5$.  In this comment we resolve this discrepancy by noting
that the anisotropy of the underlying sheet resistivities is correctly
represented by Hall bar resistance measurements but shows up
exponentially enhanced in the resistance measurements on square
samples due to simple geometric effects.  We note, however, that the
origin of this underlying resistivity anisotropy remains unknown, and
is not addressed here. 

We consider a rectangular geometry ($L_x$ by $L_y$), and assume that
the system is described by a uniform anisotropic sheet resistivity
tensor $\rho$.  In the so-called ``principle'' basis, $\rho$ is
antisymmetric.  Lack of any large observed anisotropy\cite{Lillyun}
between the resistances $R_{(x+y)(x+y)}$ and $R_{(x-y)(x-y)}$ allows
us to assume that the principle basis axes are aligned with the edges
of the sample.  To calculate the resistance we need to solve $\vec E =
\rho \vec j$, with $\nabla \cdot \vec j = 0$, and $\nabla \times \vec
E = 0$ with the boundary condition that a net current $J$ is injected
at the current source and extracted at the drain.  By writing $\vec j
= \hatn z \times \nabla \psi$, the problem is reduced to $[\rho_{yy}
\partial_x^2 + \rho_{xx} \partial_y^2] \psi = 0$ subject to the
condition that $\psi$ is a constant along the boundary except at the
source and drain where it has a step discontinuity of $\pm J$
respectively.  This equation is solved by Fourier series
methods to obtain $\psi$.  To obtain the voltage between two contacts,
we integrate the current density along the edge $\int (\hatn z \times
\nabla \psi) \cdot d\vec l$ and multiply by the longitudinal
resistivity in that direction.  To find $R_{xx}$ we place the source
and drain at the center of the faces of length $L_y$ and measure the
voltage from corner to corner along either face of length $L_x$.  We
obtain
\begin{eqnarray}
  &R_{xx}&   = \frac{4}{\pi} \sqrt{\rho_{yy} \rho_{xx}}\sum_{n = \, \mbox{\small odd}^+}
  \left[ n \sinh\left(  \sqrt{\frac{\rho_{yy}}{\rho_{xx}}} \frac{\pi n}{2}
  \frac{L_y}{L_x} \right) \right]^{-1}  \nonumber \\
&\approx&  (L_x/L_y) \rho_{xx} -
C \sqrt{\rho_{xx}\rho_{yy}} ~~~~~~~~~ \mbox{for} ~~ L_x^2 \rho_{xx} > L_y^2
  \rho_{yy} \nonumber \\
&\approx& \frac{8}{\pi} \sqrt{\rho_{yy} \rho_{xx}} \, 
  e^{-[\pi L_y/2 L_x] \sqrt{\rho_{yy}/\rho_{xx}}} ~~ \mbox{for} ~~ L_x^2 \rho_{xx} < L_y^2 \rho_{yy}
  \nonumber
\end{eqnarray}
where $C = 2 \ln 2/\pi \approx .44$. The case of $L_x \gg L_y$
corresponds to the Hall bar case, whereupon the measured resistance
$R_{xx}$ is just the actual sheet resistivity $\rho_{xx}$ times a
geometric factor $L_x/L_y$.  On the other hand, in the experimental
case of a square geometry where $L_x = L_y$, we find that an
anisotropy $\rho_{yy} > \rho_{xx}$ in the sheet resistivity results in
an exponential decrease in the measured resistance.  This decrease is
understood once we see where the current flows in a square sample.
Current flows in the $\hatn x$ direction from the source in the center
of one face to the drain in the center of the opposite face.  Most of
the current flows straight across the sample in something close to the
shortest route possible.  Only a very small amount of current ---
exponentially small in the parameter $(L_y/L_x)
\sqrt{\rho_{yy}/\rho_{xx}}$ --- extends to the edge of the sample
between the voltage contacts a distance away in the $\hatn y$
direction.  However, it is precisely this small amount of current that
determines the voltage drop measured between these two contacts.  When
the ratio $\rho_{yy}/\rho_{xx}$ is increased, the current flows more
directly in the $\hatn x$ direction, and the current between the
voltage contacts, and hence the measured voltage, decreases
exponentially.

Analogously, $R_{yy}$ is given by switching $x$ and $y$ everywhere
they occur in the above equations.  For a square sample with
$\rho_{yy} > \rho_{xx}$, we then have
$$
  R_{yy}/R_{xx} \approx (\pi/8) [
(\rho_{yy}/\rho_{xx})^{1/2} - 
  C ]
e^{(\pi/2) \sqrt{\rho_{yy}/\rho_{xx}}} 
$$
Thus an underlying anisotropy ratio of
$\rho_{yy}/\rho_{xx} \approx 7$ is sufficient to yield a measured
anisotropy ratio of 60.  It is interesting that the Hall bar
measurements yield a ratio $\rho_{yy}/\rho_{xx}
\approx 5$ which is slightly lower.  Several factors might explain
this discrepancy: (a) The two Hall bar measurements and the square
measurements are made on physically different samples (although all
are from the same wafer) and thus may not all have the same
resistivity tensors.  (b) Lack of perfect knowledge of where the
contacts are placed could slightly change the effective aspect ratio
of the sample.  (c) Misalignment of the crystal axes from the
principle axes of the resistivity tensor would change these results
(although an exponential enhancement would remain in general).
(d) Some of the conduction may be from edge state or ballistic
transport which cannot be described in terms of a simple local sheet
resistivity. (e) Large scale inhomogeneities can give spurious results
in the Hall bar measurements if the width of the Hall bar is smaller
than this disorder length scale.

I thank J. Eisenstein and M. Fogler for correcting calculational
errors in  previous versions of this comment.

\vspace*{5pt}

\noindent Steven H. Simon \\
Lucent Technologies, Bell Labs \\
Murray Hill, NJ 07974


\end{document}